\documentstyle[12pt,epsfig,amssymb]{article}
\begin{document}
\setlength{\parskip}{0.45cm}
\setlength{\baselineskip}{0.75cm}
\begin{titlepage}
\begin{flushright}
DO-TH 96/10 \\ RAL-TR-96-033 \\ May 1996
\end{flushright}
\vspace{1.0cm}
\begin{center}
\Large
{\bf Photoproduction of Jets and Heavy Flavors}

\vspace{0.1cm}
{\bf at Future Polarized {\em{ep}} - Colliders}

\vspace{1.5cm}
\large
M.\ Stratmann\\
\vspace{0.5cm}
\normalsize
Universit\"{a}t Dortmund, Institut f\"{u}r Physik, \\
\vspace{0.1cm}
D-44221 Dortmund, Germany \\
\vspace{1.2cm}
\large
W. Vogelsang \\
\vspace{0.5cm}
\normalsize
Rutherford Appleton Laboratory \\
\vspace{0.1cm}
Chilton, Didcot, Oxon OX11 0QX, England \\
\vspace{2cm}
{\bf Abstract}
\vspace{-0.3cm}
\end{center}
We study photoproduction of jets and heavy flavors 
in a polarized collider mode
of HERA and in polarized $ep$ collisions at $\sqrt{S}\approx 30$ GeV.
We examine the sensitivity of the cross sections and their asymmetries 
to the proton's polarized gluon distribution and to the completely unknown 
parton distributions of longitudinally polarized photons.
\end{titlepage}
\section{Introduction}
The last few years have brought much new experimental information on the
spin structure of the nucleon via measurements \cite{data} of the
spin asymmetries $A_1^N$ ($N=p,n,d$) in longitudinally polarized
deep-inelastic scattering (DIS) of leptons off polarized nucleon targets.
Recent theoretical leading order (LO) \cite{grvp,grsv,gs} and
next-to-leading order (NLO)  \cite{grsv,bfr,gs} analyses of the data
sets demonstrate, however, that these are not sufficient to accurately
extract the spin-dependent quark ($\Delta q = q^{\uparrow}-q^{\downarrow}$) 
and gluon ($\Delta g=g^{\uparrow}-g^{\downarrow}$) densities of the nucleon. 
This is true in particular for $\Delta g(x,Q^2)$ since it enters DIS in LO
only via the $Q^2$-dependence of $g_1$ (or $A_1$) which could not yet be 
accurately studied experimentally. As a result of this, it turns out 
\cite{grsv,gs} that the $x$-shape of 
$\Delta g$ seems to be hardly constrained 
at all by the DIS data, even though a tendency towards a sizeable positive 
{\em total} gluon polarization, $\int_0^1 \Delta g(x,Q^2=4 \; \mbox{GeV}^2) 
dx \gtrsim 1$, was found 
\cite{grsv,bfr,gs}. Clearly, the measurement of $\Delta g$ is one of the 
most interesting challenges for future spin physics experiments.

Among the various conceivable options for future HERA upgrades is the
idea to longitudinally polarize its proton beam \cite{barber} which, 
when combined with the already operative longitudinally polarized electron 
(positron) beam, results in a polarized version of the usual HERA collider 
with $\sqrt{S}=298$ GeV. A typical conservative value for the integrated 
luminosity in this case should be 100 $\mbox{pb}^{-1}$, but higher 
luminosities, up to 1000 $\mbox{pb}^{-1}$ might not be inconceivable for 
future HERA upgrades. HERA has already been very successful in pinning down
the proton's unpolarized gluon distribution $g(x,Q^2)$. Apart from
exploring the unpolarized DIS structure function $F_2$ over a wide
range in $x$ and $Q^2$ \cite{f2} which indirectly constrains $g(x,Q^2)$ 
in global fits via scaling violations \cite{scavio,f2}, 
also processes have been studied which have contributions from
$g(x,Q^2)$ already in the lowest order. Among these are (di)jet and
heavy flavor production. Since events at HERA are concentrated in the
region $Q^2 \rightarrow 0$, the processes have first and most
accurately been studied in photoproduction [9-14].
As is well-known, in this case the (quasi-real) photon will not only
interact in a direct ('point-like') way, but can also be resolved into 
its hadronic structure. HERA photoproduction experiments like [9-12]
have not merely established evidence for
the existence of such a resolved contribution, but have also been precise
enough to improve our knowledge about the parton distributions, 
$f^{\gamma}$, of the photon. 
Here they have provided information complementary 
to the results for $F_2^{\gamma}$ obtained in various $e^+e^-$ experiments,
by constraining the photonic gluon distribution \cite{jet2h1}.
More recently, the production of two jets in DIS events ($Q^2
\neq 0$) has been used for a first direct measurement of $g(x,Q^2)$
\cite{jetq2}, and first results for the charm contribution, $F_2^c (x,Q^2)$,
to $F_2$ have been presented \cite{f2ch}.

Given the success of such unpolarized experiments at HERA, it seems
most promising to closely examine the same processes for the situation with
longitudinally polarized beams with regard to their sensitivity 
to $\Delta g$, which is the main purpose of this paper. Here we will focus 
on the {\em photo}production of open charm and jets.
Firstly, as mentioned above, photoproduction experiments will yield 
the largest event rates and are 
thus expected to be more accurate. Furthermore, 
they may in principle allow to not only determine the parton, in 
particular gluon, content of the polarized {\em proton}, but also 
that of the longitudinally polarized {\em photon} which is completely
unknown so far. Since, e.g., a measurement of the photon's spin-dependent
structure function $g_1^{\gamma}$ in polarized $e^+ e^-$ collisions is not
planned in the near future, HERA could play a unique role here, even if it
should only succeed in establishing the very {\em existence} of a resolved
contribution to polarized photon-proton reactions. We emphasize at this point 
that the role of this contribution to photoproduction processes with 
polarized beams at HERA has never been investigated before: In \cite{kun}
polarized photoproduction of dijets at HERA was suggested for the first time
as a possible tool for measuring $\Delta g$, but the expected cross section
and asymmetry were only roughly estimated, based on the single process
$\gamma g\rightarrow q\bar{q}$ with the rather optimistic assumption
$\Delta g/g=0.5$, and on neglecting the resolved contribution to the
cross section. Very recently, a study of 
open-charm photoproduction at polarized HERA was presented \cite{fr}. 
Again, the contribution to the cross section arising from resolved polarized 
photons was neglected. Even though it will turn out that for charm production 
this is a fairly good approximation in most cases, it clearly needs to be 
checked. We also believe that the issue of the sensitivity of the charm cross 
section to $\Delta g$ was not thoroughly discussed in \cite{fr}. 

We note that there are also ideas 
for a high-luminosity polarized collider with 
(typically) 5 GeV electrons on 50 GeV protons at the GSI \cite{gsi}. Such 
energies will be too low for jet physics, but are certainly appropriate for 
charm production. We will therefore extend our charm predictions 
also to this situation. As we will see, 
some results look more promising for lower energies since the corresponding
cross section asymmetries are larger than in the HERA situation.
We remark that the process $\gamma p \rightarrow c\bar{c} X$ with polarized
photons and protons at (comparably low) fixed target energies has 
originally been suggested in the literature in \cite{gr} and further studied
in [22-25]. In fact, a measurement of the spin asymmetry for 
the total charm photoproduction cross section is planned in a fixed target 
$\mu p$ - experiment with $\sqrt{S}\approx 14$ GeV by the COMPASS 
collaboration \cite{compass}. We will essentially 
update and/or extend the previous studies [21-25] to the
GSI $ep$ - situation by using more up-to-date sets of polarized parton
distributions covering the whole range of allowed $\Delta g$,
and also by providing studies of transverse momentum and 
rapidity distributions of the produced charm quarks, which should be
accessible with high luminosity at the GSI machine. 

The paper is organized as follows: In section 2 we collect the necessary
ingredients for our calculations, like the parton distributions of the 
proton and photon we use. Section 3 is devoted to charm production, 
which we study in terms of the fully inclusive charm contribution,
$g_1^c$, to the polarized DIS structure function $g_1$, but mainly 
as polarized open-charm photoproduction at HERA and the GSI. In section 4 we 
examine polarized photoproduction of (di)jets. Section 5 contains the
conclusions. 
%
%
\section{Polarized Parton Distributions of the Proton and the Photon}
As stated in the introduction, theoretical analyses of polarized 
DIS which take into account all or most data sets \cite{data}, have been 
published recently \cite{grsv,bfr,gs}. For the first time, these studies 
could even go to NLO of QCD, since the NLO framework for polarized
DIS had become complete due to the calculation of the spin-dependent 
NLO $Q^2$-evolution kernels \cite{MvN}. Since, however, the NLO
corrections to polarized charm or jet production are not yet known,
we have to stick to LO calculations throughout this work, which 
implies use of LO parton distributions\footnote{Note that the recent 
study \cite{fr} uses polarized parton distributions evolved in NLO in the 
LO calculation of the charm cross section for HERA. While this may 
sufficiently serve the purpose of numerically checking the sensitivity of the 
cross section to $\Delta g$, it is theoretically inconsistent since it 
introduces a factorization scheme dependence to the cross section.}. 
Fortunately, the studies 
\cite{grsv,gs} also provide LO sets of the proton's polarized parton
distributions which give an accurate description of all presently 
available DIS data. Both papers give various LO sets which mainly differ 
in the $x$-shape of the polarized gluon 
distribution, which turns out to be hardly constrained by DIS.
For definiteness, we will choose the LO 'valence' set of the 'radiative 
parton model analysis' \cite{grsv}, which corresponds to the best-fit 
result of that paper, along with two other sets of \cite{grsv} which 
are based on very different assumptions about the polarized gluon 
distribution at the low input scale $\mu$ of \cite{grsv}: One set assumes 
$\Delta g (x,\mu^2) = g(x,\mu^2)$, which is the maximally allowed gluon 
input distribution obeying the fundamental positivity constraints
\begin{equation}  \label{poscon}
|\Delta f (x,\mu^{2})|\leq f (x,\mu^{2})  
\end{equation}
($f=q,g$), where the $f(x,\mu^2)$ are the unpolarized LO GRV \cite{grv} 
input distributions. 
The other set adopts $\Delta g(x,\mu^2)=0$. It turns out that the
sets A,B of \cite{gs} (GS) have gluon distributions similar to 
the above maximal one of \cite{grsv}, only the gluon of set C of \cite{gs}
is qualitatively different since it has a large negative polarization 
at large $x$. We will therefore also use this set in our calculations.
For illustration, we show in Fig.~1 the gluon distributions  
of the four different sets of parton distributions we will use, taking a 
typical scale $Q^2=10$ GeV$^2$. 
Keeping in mind that all four sets provide very 
good descriptions of all present polarized DIS data \cite{data}, 
it becomes obvious that the data indeed do not seem to be able to 
significantly constrain the $x$-shape of $\Delta g(x,Q^2)$.

As we pointed out in the introduction, we will mainly consider 
photoproduction in this paper. In this case the electron just
serves as a source of quasi-real photons which are radiated according 
to the Weizs\"{a}cker-Williams spectrum \cite{ww}. The photons can then 
interact either directly or via their partonic structure ('resolved' 
contribution). In the case of longitudinally polarized electrons, the 
resulting photon will be longitudinally (more precisely, circularly)
polarized and, in the resolved case, the {\em polarized}
parton distributions of the photon enter the calculations. Thus one
can define the effective polarized parton densities at the scale $M$ 
in the longitudinally polarized electron via
\begin{equation}  \label{elec}
\Delta f^e (x_e,M^2) = \int_{x_e}^1 \frac{dy}{y} \Delta P_{\gamma/e} (y)
\Delta f^{\gamma} (x_{\gamma}=\frac{x_e}{y},M^2) \; 
\end{equation}
($f=q,g$) where $\Delta P_{\gamma/e}$ is the 
polarized Weizs\"{a}cker-Williams
spectrum and $\Delta f^{\gamma} (x_{\gamma},M^2)$ are the polarized photon
structure functions with the additional definition $\Delta f^{\gamma} 
(x_{\gamma},M^2) \equiv \delta (1-x_{\gamma})$ for the direct ('unresolved') 
case. 

The parton distributions of polarized photons are completely unmeasured 
so far, such that models for the $\Delta f^{\gamma} (x,M^2)$ have 
to be invoked. In the unpolarized case, a phenomenologically very successful
prediction for the unpolarized photon structure functions $f^{\gamma}$
has emerged within the radiative parton model, where \cite{grvg} a VMD 
valence-like structure at a low resolution scale $\mu$ was imposed as the 
input boundary condition, assuming that at this scale the photon entirely 
behaves like a vector meson, i.e., that its parton content is proportional 
to that of the $\rho$-meson. Since nothing is known experimentally about
the latter, the parton densities of the neutral pion as determined in
a previous study \cite{grvpi} were used instead which are expected not 
to be too dissimilar from those of the $\rho$. In \cite{gvg,gsvg} a similar 
approach was adopted for the polarized case where, however, it is   
obviously impossible to uniquely fix the VMD input distributions 
$\Delta f^{\gamma}(x,\mu^2)$ in this way.
Therefore, to obtain a realistic estimate for the theoretical uncertainties 
in the polarized photon structure functions coming from the unknown
hadronic input, two very different scenarios were considered in 
\cite{gvg,gsvg} with 'maximal' ($\Delta f^{\gamma}(x,\mu^2)=
f^{\gamma}(x,\mu^2)$) and 'minimal' ($\Delta f^{\gamma}(x,\mu^2)=0$) 
saturation of the fundamental positivity constraints (\ref{poscon}).
The results of these two extreme approaches are presented in Fig.~2 in terms 
of the photonic parton asymmetries $A_f^{\gamma} \equiv \Delta f^{\gamma}/
f^{\gamma}$, evolved to $Q^2=30$ GeV$^2$ in LO. An ideal aim of 
measurements in a polarized collider mode of HERA would of course be to 
determine the $\Delta f^{\gamma}$ and to see which ansatz for the 
hadronic component is more realistic. The sets presented in Fig.~2, which we 
will use in what follows, should in any case be sufficient to study 
the sensitivity of the various cross sections to the $\Delta f^{\gamma}$,
but also to see in how far they influence a determination of $\Delta g$.

We still have to specify the polarized Weizs\"{a}cker-Williams spectrum 
which we will use in our calculations:
\begin{equation}  \label{weiz}
\Delta P_{\gamma/e} (y) = \frac{\alpha_{em}}{2\pi} \left[ 
\frac{1-(1-y)^2}{y} \right] \ln \frac{Q^2_{max} (1-y)}{m_e^2 y^2} \; ,
\end{equation}
where $m_e$ is the electron mass. For the time being, it seems most 
sensible to follow as closely as possible the analyses successfully 
performed in the unpolarized case, which implies to introduce the same 
kinematical cuts. As in \cite{jet1ph,jet2ph,chph,kramer} we will use
an upper cut\footnote{In H1 analyses of HERA photoproduction
data \cite{jet1h1,jet2h1,chh1}
the cut $Q^2_{max}=0.01$ GeV$^2$ is used along with slightly
different $y$-cuts as compared to the corresponding 
ZEUS measurements \cite{jet1ph,jet2ph,chph}, which leads to
smaller rates.} 
$Q^2_{max}=4$ GeV$^2$, and the $y$-cuts $0.2 \leq y \leq 0.85$ 
(for charm and single-jet \cite{jet1ph} production) 
and $0.2 \leq y \leq 0.8$ (for dijet production, \cite{jet2ph}) 
will be imposed. We note that a larger value for the lower limit, $y_{min}$, 
of the allowed $y$-interval would enhance the yield of polarized photons 
relative to that of unpolarized ones since $\Delta P_{\gamma/e}(y)/
P_{\gamma/e}(y)$, where $P_{\gamma/e}$ is the unpolarized 
Weizs\"{a}cker-Williams spectrum given by 
\begin{equation}  \label{weizunp}
P_{\gamma/e} (y) = \frac{\alpha_{em}}{2\pi} \left[ 
\frac{1+(1-y)^2}{y} \right] \ln \frac{Q^2_{max} (1-y)}{m_e^2 y^2} \; ,
\end{equation}
is suppressed for small $y$. On the other hand, increasing $y_{min}$ 
would be at the expense of reducing the individual polarized and
unpolarized rates.

We finally note that in what follows a polarized cross section will always
be defined as
\begin{equation} 
\Delta \sigma \equiv \frac{1}{2} \left( \sigma (++)-\sigma (+-) \right) \; ,
\end{equation}
the signs denoting the helicities of the scattering particles. The
corresponding unpolarized cross section is given by  
\begin{equation} 
\sigma \equiv \frac{1}{2} \left( \sigma (++)+\sigma (+-) \right) \; ,
\end{equation}
and the cross section asymmetry is $A\equiv \Delta \sigma/\sigma$.
Whenever calculating an asymmetry $A$, we will 
use the LO GRV parton distributions for the proton \cite{grv} and the 
photon \cite{grvg} to calculate the unpolarized cross section, which 
guarantees satisfaction of the positivity constraints (\ref{poscon}).
For consistency, we will employ the LO expression for the strong coupling 
$\alpha_s$, 
\begin{equation}
\alpha_s (Q^2) = \frac{12 \pi}{(33-2 f) \ln (Q^2/\Lambda_{QCD}^2)} \; ,
\end{equation}
where $f$ is the number of active flavors. All unpolarized \cite{grv,grvg} and 
polarized \cite{grsv,gs,gvg,gsvg} parton distributions we employ have been set 
up using $\Lambda_{QCD}^{(f=4)}=200$ MeV, which eliminates possible 
mismatches.
%
%
\section{Charm Production at HERA and the GSI}
To begin with, we study the charm contribution, $g_1^c$, to the 
spin-dependent DIS structure function $g_1$. The motivation for this 
is that $g_1^c$ is expected to be driven by the polarized photon-gluon
fusion (PGF) subprocess\footnote{This approach implies not to treat charm as 
a parton of the proton which is realized in the sets of unpolarized \cite{grv} 
and polarized \cite{grsv,gs} parton distributions we use.} 
$\gamma^{*}g\rightarrow c\bar{c}$. Furthermore, 
recently first HERA results for the unpolarized $F_2^c$ have been reported 
\cite{f2ch}, such that a measurement of $g_1^c$ in a future high-luminosity
experiment with polarized beams does not seem completely unrealistic. 
The results of \cite{f2ch} also indicate that PGF is indeed the 
correct mechanism for charm production. In the polarized case, its 
contribution to $g_1$ is given by \cite{watson,grvalt}
\begin{equation}   \label{eq1}
g_1^c (x,Q^2) = \frac{\alpha_s (M^2)}{9\pi} \int_{ax}^1 \frac{dy}{y}
\Delta h (\frac{x}{y},Q^2) \Delta g (y,M^2) \; ,
\end{equation}
where 
\begin{equation}  \label{dh}
\Delta h (z,Q^2) = (2z-1) \ln \frac{1+\beta}{1-\beta} + \beta (3-4 z)
\end{equation}
with $\beta^2 (z) = 1-4 m_c^2 z/Q^2 (1-z)$. In (\ref{eq1}), $a=1+4 m_c^2/Q^2$
and $M$ is some mass scale for which we will use $M=2 m_c$ (with the charm 
mass $m_c=1.5$ GeV) which was shown \cite{grs} to lead to a good perturbative 
stability of predictions for the unpolarized $F_2^c$.  

Fig.~3 shows our results for $g_1^c (x,Q^2)$ and the charm asymmetry
$A_1^c \equiv g_1^c/F_1^c$ (with $F_1^c = (F_2^c-F_L^c)/2x$ calculated 
according to the unpolarized PGF process as given, e.g., in \cite{grs}) 
at $Q^2=10$ GeV$^2$ for the four different
gluon distributions introduced in the last section. As becomes obvious,
the results for the gluon distributions of \cite{grsv} nicely reflect the
relative size of the distributions in the range $x > 0.003$. However, 
the gluon distribution C of \cite{gs} gives a result which is
rather surprising at a first glance for small values of $x$, the reason
being the convolution of the oscillating gluon distribution (see Fig.~1) with 
the subprocess cross section $\Delta h(z,Q^2)$ (see eq.(\ref{dh})) which also 
has a zero. It becomes clear from these examples that data points at {\em 
several different} values of $x$ would be 
needed in order to really pin down the shape of $\Delta g$. We note that 
the dashed line in Fig.~3a at $x\approx 0.005$ corresponds to about 
$10\%$ of the full $g_1$ which implies that a fairly accurate 
measurement would be required. Fig.~3b also shows that the deep-inelastic 
charm asymmetry $A_1^c$ is of the order of $5\%$ or less in this $x$-region. 
It becomes much larger at larger values of $x$ where, however, the individual 
$g_1^c$ and $F_1^c$ (or $F_2^c$) rapidly decrease as a result of the 
threshold condition $\beta^2 \geq 0$. 

We now turn to the case of photoproduction of charm. For illustration, 
let us first consider the total cross section. In the unpolarized case 
it has been possible to extract the total cross section for $\gamma p 
\rightarrow c\bar{c}$ from the fixed target \cite{oldch} and HERA 
\cite{chph,chh1} lepton-nucleon data, i.e., the open-charm cross section for a 
fixed photon energy without the smearing from the Weizs\"{a}cker-Williams 
spectrum. To LO, the corresponding polarized cross section is given by 
\begin{equation}   \label{wqctot}
\Delta \sigma^c (S_{\gamma p}) =\sum_{f^{\gamma},f^p} 
\int_{4m_c^2/S_{\gamma p}}^1
d x_{\gamma} \int_{4m_c^2/x_{\gamma}S_{\gamma p}}^1 d x_p
\Delta f^{\gamma} (x_{\gamma},M^2) \Delta f^p (x_p,M^2)
\Delta \hat{\sigma}^c (\hat{s},M^2) \; .
\end{equation}
where the $\Delta f^p$ stand for the polarized parton 
distributions of the proton and $\hat{s} \equiv x_{\gamma} 
x_p S_{\gamma p}$. For the direct (unresolved) contribution, $\Delta 
f^{\gamma} (x_{\gamma},M^2) = \delta (1-x_{\gamma})$ is to be understood. 
In this case, the contributing subprocess is again only PGF, 
$\gamma g \rightarrow c\bar{c}$, the spin-dependent total LO subprocess cross 
section $\Delta \hat{\sigma}^c (\hat{s})$ for which can be found in 
\cite{gr,grvalt}. In the resolved case, the processes 
$gg \rightarrow c\bar{c}$ and $q\bar{q} 
\rightarrow c\bar{c}$ contribute; their cross sections have been calculated 
in \cite{cont}. Needless to say that we can obtain the corresponding 
unpolarized LO charm cross section $\sigma^c (S_{\gamma p})$ by using 
LO unpolarized parton distributions and subprocess cross sections 
(as calculated in \cite{chunp}) in (\ref{wqctot}). We note that recent 
HERA results for $\sigma^c (S_{\gamma p})$ are well-described by
LO calculations based on the PGF process and use of standard unpolarized 
LO distributions such as \cite{grv}.

Fig.~4 shows the result for the asymmetry $\Delta \sigma^c/\sigma^c$ for the 
four different sets of polarized parton distributions, where we have again 
used the scale $M=2 m_c$. The resolved contribution to the cross section is 
rather small in the unpolarized case. 
For the polarized case, we have calculated
it using the 'maximally' saturated set for the polarized photon structure
functions, which should roughly provide the maximally possible 
background from resolved photons. The contribution, which was neglected in
\cite{fr}, turns out to be non-negligible only for 
large $\sqrt{S_{\gamma p}}$, where it can be as large as about 1/3 the direct
contribution but with opposite sign. As becomes obvious from Fig.~4, 
the asymmetry becomes very small \cite{fr} towards the HERA region at larger 
$\sqrt{S_{\gamma p}} \sim 200$ GeV. 
One reason for this is the oscillation of the 
polarized subprocess cross section for the direct part, combined with 
cancellations between the direct and the resolved parts. 
More importantly, as seen from eq.(\ref{wqctot}),
the larger $S_{\gamma p}$ becomes, the smaller are the $x_{p,\gamma}$ - values 
probed, such that the rapid rise of the unpolarized parton distributions
strongly suppresses the asymmetry. The measurement of the {\em total} charm
cross section asymmetry in $\gamma p \rightarrow c\bar{c}$ therefore seems 
rather more feasible at smaller energies, $\sqrt{S_{\gamma p}} 
\lesssim 20$ GeV, i.e., in the region where measurements 
at the GSI and in the COMPASS experiment \cite{compass} would be performed
and where also the unknown resolved contribution to the cross section is 
negligible. Here the differences between the various gluon distributions 
show up rather strongly, even though measurements at various different 
$\sqrt{S_{\gamma p}}$ would be needed to decide between the gluons. 
For completeness we present in Table 1 some numbers for the total cross 
sections. 

\noindent
\begin{table}[htb]
%
\hbox to \textwidth {\hss
\begin{tabular}{c|c|c||c|c||c|c||c|c|}
&\multicolumn{2}{c||}{fitted gluon}&
 \multicolumn{2}{c||}{$\Delta g = g$ input} &
 \multicolumn{2}{c||}{$\Delta g = 0$ input} &
 \multicolumn{2}{c|}{GS C} \\ \cline{2-9} 
$\sqrt{S_{\gamma p}}$&
\multicolumn{1}{c}{direct} &
\multicolumn{1}{c||}{resolved} &
\multicolumn{1}{c}{direct} &
\multicolumn{1}{c||}{resolved} &  
\multicolumn{1}{c}{direct} &
\multicolumn{1}{c||}{resolved} &
\multicolumn{1}{c}{direct} &
\multicolumn{1}{c|}{resolved} \\
$\mbox{[GeV]}$& 
\multicolumn{1}{c}{$\mbox{[nb]}$} &
\multicolumn{1}{c||}{$\mbox{[nb]}$} &
\multicolumn{1}{c}{$\mbox{[nb]}$} &
\multicolumn{1}{c||}{$\mbox{[nb]}$} &  
\multicolumn{1}{c}{$\mbox{[nb]}$} &
\multicolumn{1}{c||}{$\mbox{[nb]}$} &
\multicolumn{1}{c}{$\mbox{[nb]}$} &
\multicolumn{1}{c|}{$\mbox{[nb]}$} \\
\hline\hline
10&10.2&-0.72&23.0&-0.63&3.46&-0.74&-3.48&-0.70 \\ \hline
20&13.9&-0.29&23.2&0.33&3.26&-0.56&19.6&-0.70 \\ \hline
30&9.32&0.39&13.7&1.48&1.52&-0.22&25.0&-0.35 \\ \hline
50&2.07&1.30&1.15&2.94&-0.53&0.18&17.4&0.61 \\ \hline
100&-4.76&2.09&-9.71&4.00&-1.90&0.46&0.58&2.27 \\ \hline
200&-7.00&2.06&-12.8&3.60&-1.96&0.44&-8.79&3.23 \\ \hline
300&-6.84&1.71&-12.2&2.84&-1.70&0.33&-10.4&3.21 \\ \hline
\end{tabular}
\hss}
\caption{Total cross sections for charm photoproduction in polarized
$\gamma p$ collisions.}
\end{table}

\hspace*{0cm}From our observations for 
HERA-energies it follows that it could be more 
promising to study distributions of the cross section in the 
transverse momentum or the pseudorapidity of the charm quark in order to 
cut out the contributions from very small $x_{p,\gamma}$. We will now 
include the Weizs\"{a}cker-Williams
spectrum since tagging of the electron, needed for the extraction 
of the cross section at fixed photon energy, will probably reduce the cross 
section too strongly. The polarized LO cross section for 
producing a charm quark with transverse momentum $p_T$ and 
cms-pseudorapidity $\eta$ then reads 
\begin{equation} \label{wqc}
\frac{d^2 \Delta \sigma^c}{dp_T d\eta} = 2 p_T 
\sum_{f^e,f^p} \int_{\frac{\rho e^{-\eta}}
{1-\rho e^{\eta}}}^1 d x_e x_e \Delta f^e (x_e,M^2) x_p \Delta f^p (x_p,M^2)
\frac{1}{x_e - \rho e^{-\eta}} \frac{d\Delta \hat{\sigma}}{d\hat{t}} \; ,
\end{equation}
where $\rho \equiv m_T/\sqrt{S}$ with $m_T \equiv \sqrt{p_T^2+m_c^2}$, and 
$$x_p \equiv \frac{x_e \rho e^{\eta}}{x_e - \rho e^{-\eta}} \; . $$
We note that both the HERA and the GSI kinematics are asymmetric since
$E_p \neq E_e$. The cross section can be transformed to the more relevant
laboratory frames by a simple boost which implies 
$$\eta \equiv \eta_{cms} = \eta_{LAB} -\frac{1}{2} \ln \frac{E_p}{E_e} \; ,$$
where we have, as usual, counted positive pseudorapidity in the proton
forward direction.
The polarized 'electronic' parton distributions $\Delta f^e (x_e,M^2)$ 
($f=\gamma,q,g$) in (\ref{wqc}) are as defined in eq.(\ref{elec}). 
The spin-dependent differential LO subprocess cross sections 
$d\Delta \hat{\sigma}/d\hat{t}$ for the resolved processes 
$gg \rightarrow c\bar{c}$ and $q\bar{q} \rightarrow c\bar{c}$ with $m_c 
\neq 0$ can again be found in \cite{cont}. The polarized cross section for 
the direct subprocess $\gamma g \rightarrow c\bar{c}$ is readily obtained 
from that for $gg\rightarrow c\bar{c}$ by dropping the non-abelian part and 
multiplying by $2 N_C e_c^2 \alpha_{em}/\alpha_s$ where $e_c=2/3$. For the
factorization/renormalization scale in (\ref{wqc}) we choose $M=m_T/2$; 
we will comment on the scale dependence of the results at the end of this 
section.

Fig.~5 shows our results for the four different sets of polarized parton 
distributions for the HERA case with $E_p=820$ GeV and $E_e=27$ GeV.  
Fig.~5a displays the $p_T$-dependence of the cross section, where we have 
integrated over $-1\leq \eta_{LAB} \leq 2$. The resolved contribution to the 
cross section has been included, calculated with the 'maximally' saturated 
set of polarized photon structure functions. It is shown individually for
the 'fitted $\Delta g$'-set of polarized proton distributions by the lower 
solid line in Fig.~5a.
Comparison of the two solid lines in Fig.~5a shows that the resolved
contribution is negligibly small\footnote{We note that the neglect 
of the resolved contribution in \cite{fr} was therefore justified in this
case.} in this case unless $p_T$ becomes very
small. The cross section in Fig.~5a should be large enough to be 
measurable even at $p_T \approx 15$ GeV if high luminosities can be attained.
Fig.5b shows the asymmetries corresponding to Fig.~5a. It becomes obvious 
that they are much larger than for the total cross section 
if one goes to $p_T$ of about 10-20 GeV, which is in agreement 
with the corresponding findings of \cite{fr}. Furthermore, one sees that 
the asymmetries are strongly sensitive to the size {\em and} shape of the 
polarized gluon distribution used. Similar statements are true for the 
$\eta_{LAB}$-distributions shown in Figs.~5c,d, where $p_T$ has been 
integrated over $p_T>8$ GeV in order to increase the number of events. 
Even here the resolved contribution remains small, although it becomes 
more important towards large positive values of $\eta_{LAB}$. 

Fig.~6 shows similar results for the GSI situation with
$E_p=50$ GeV and $E_e=5$ GeV. For Figs.~6a,b, $\eta_{LAB}$ has been   
integrated over $-1\leq \eta_{LAB} \leq 1$, and for Figs.~6c,d all
events with $p_T>3$ GeV have been collected. Again we find possibly measurable 
cross sections with very promising asymmetries, reaching up to 
about $-40 \%$ for the set with the largest gluon
distribution. This time, since fairly large values of $x_p$ are probed,
the negative large-$x$ polarization of gluon C of \cite{gs} shows up 
prominently as an asymmetry of different sign. We note that 
all polarized cross sections in Figs.~5,6 are {\em negative} (apart 
from the one for GS C). This is surprising at first sight for the 
results at lower energy in Fig.~6 since the asymmetry for the {\em total}
charm photoproduction cross section in Fig.~4 and Table~1 was {\em positive} 
for small $\sqrt{S_{\gamma p}}$. It turns out that the cross sections 
in Figs.~5,6a actually change sign at small $p_T$ and obtain a large
positive contribution from the region $p_T\lesssim 
m_c$ which, when integrating over $p_T$, compensates for the negative
contribution at large-$p_T$ shown in Figs.~5,6a. This feature once more 
demonstrates that in the polarized case distributions in $p_T$ or $\eta$ can 
be more expedient than the total cross section.

We finally briefly address the theoretical uncertainties of our results in
in Figs.~5,6 related to the dependence of the cross sections and asymmetries 
on the renormalization/factorization scale $M$. Since all our calculations
could be performed in LO only, this is a particularly important issue. 
We have therefore recalculated the results in Figs.~5,6, now using the 
scale $M=m_T$. As a result, it turns out that the cross 
sections in Fig.~5a are subject to changes of about $10\%$ at $p_T<15$ GeV,
and of as much as $20-25\%$ at larger $p_T$. Changes of in most cases below
$10\%$ are found for the $\eta_{LAB}$-curves in Fig.~5c. Similar statements 
with generally slightly larger scale uncertainties apply to our results for 
lower energies in Figs.~6a,c. In contrast to this (not unexpected) fairly 
strong scale dependence of the polarized cross sections, the {\em asymmetries}, 
which will be the quantities actually measured, are very insensitive to 
scale changes, deviating usually by not more than a few percent from the 
values shown in Figs.~5,6 b,d for all relevant $p_T$ and  $\eta_{LAB}$. 
This finding seems important in two respects: Firstly, 
it warrants the genuine sensitivity of the asymmetry to $\Delta g$, implying
that despite the sizeable scale dependence of the cross section it still seems
a reasonable and safe procedure to compare LO theoretical predictions for the 
asymmetry with future data and to extract $\Delta g$ from such a comparison.
Secondly, it sheds light on the possible role of NLO corrections to 
our results, suggesting that such corrections might be sizeable for the 
cross sections, but less important for the asymmetry. This conjecture can only 
be confirmed or disproved once the NLO corrections to the polarized 
charm cross sections are known, which would be desirable for the future. 
We note, however, that previous experience with the spin asymmetry for 
prompt-photon production in hadronic collisions and the NLO corrections 
to it \cite{congam} supports this view.
%
%
\section{Photoproduction of Jets at HERA}
In this section we study photoproduction of jets. We restrict ourselves
to the HERA situation since the energy of the GSI-collider would probably 
not be sufficient for jet physics. 

The generic cross section formula for the production of a single jet 
with transverse momentum $p_T$ and rapidity $\eta$ is similar to
that in (\ref{wqc}), the sum now running over all properly symmetrized 
$2\rightarrow 2$ subprocesses for the direct ($\gamma b\rightarrow cd$) 
and resolved ($ab\rightarrow cd$) cases. When only light flavors are involved 
one uses $m_c=0$ in (\ref{wqc}), and the corresponding differential 
helicity-dependent LO subprocess cross sections can be found in \cite{bab}. 
A sensible way of estimating the contribution to the cross section coming 
from charm quarks would be to use the (properly symmetrized) matrix elements 
of \cite{cont,chunp} already employed for Figs.~5,6 in the previous section, 
which fully take into account the charm mass and threshold effects. 
We found that for the values of $p_T$ we consider in the following the effects
of the finite charm mass can be safely neglected due to $m_c^2/\hat{s} \ll 1$, 
such that in all following predictions we will deal with the charm 
contribution to the cross section by including charm as a massless 
{\em{final}} state 
particle (see footnote 3) via the subprocesses $\gamma g \rightarrow c\bar{c}$ 
(for the direct part) and $gg \rightarrow c\bar{c}$, $q\bar{q} 
\rightarrow c\bar{c}$ (for the resolved part). In all following applications 
we will use the renormalization/factorization scale $M=p_T$. We have again 
found that the scale dependence of the asymmetries is rather weak as 
compared to that of the cross sections.

Fig.~7 shows our results for the single-inclusive jet cross section and its 
asymmetry as a function of $p_T$ and integrated over $-1 \leq \eta_{LAB} 
\leq 2$ for the four sets of the polarized proton's parton distributions.
For Figs.~7a,b we have used the 'maximally' saturated set of polarized 
photonic parton densities, whereas Figs.~7c,d correspond to the 
'minimally' saturated set. Figs.~7a,c show that the polarized cross
section is quite substantial for $p_T$ not too large, $p_T \lesssim
15$ GeV. It is obviously also very sensitive
to the polarized gluon distribution of the proton. At the same time, 
however, the resolved contribution to the cross section strongly dominates
in the small-$p_T$ region, as can be seen from comparison of the 
results in Figs.~7a,c or 7b,d. Keeping in mind that the 'true' set of 
polarized photon structure functions may well lie somewhere between the two 
extreme sets we use, this implies that, unless an experimental
distinction between resolved and direct contributions can be achieved,
it will hardly be possible to make a clear statement about the size 
of $\Delta g$ and/or the polarized photonic parton distributions from 
a measurement of the jet cross section at these values of $p_T$.
Furthermore, as can be seen from Figs.~7b,d, the asymmetry is very 
small below $p_T=15$ GeV, which is a result of the fact that the parton
distributions are mainly probed at small values of $x$, and of 
the dominance of the resolved piece (with its many contributing subprocesses)
also in the unpolarized case, consequently further diluting the asymmetry. 
As far as a clear-cut 
sensitivity to $\Delta g$ is concerned, the situation improves at
larger $p_T$. Here the direct contribution clearly dominates,
and the asymmetries become larger. On the other hand, the polarized cross
section is very small already at $p_T\approx 25$ GeV even for the set with the
largest $\Delta g$. 

It appears more promising to study the $\eta_{LAB}$-distribution of the 
cross section and the asymmetry. The reason for this is that for negative 
$\eta_{LAB}$ the main contributions are expected to come from the region
$x_{\gamma} \rightarrow 1$ and thus mostly from the direct piece
at $x_{\gamma}=1$. To investigate this, Fig.~8 repeats the analysis
presented in Fig.~7, but now as a function of $\eta_{LAB}$ with 
$p_T$ integrated over $p_T>8$ GeV. Comparison of Figs.~8a,c or
8b,d (which differ in the polarized photon structure functions used) shows 
that indeed the direct contribution clearly dominates for $\eta_{LAB} \leq
-0.5$, where also differences between the polarized gluon distributions 
of the proton show up clearly. Furthermore, the cross sections are  
generally large in this region with asymmetries of a few percents. At positive 
$\eta_{LAB}$, we find the same picture as for the $p_T$-dependence of the 
cross section in Fig.~7 at small $p_T$: The cross section is dominated 
by the resolved contribution and is therefore sensitive to
both the parton content of the polarized proton {\em and} the photon.
It turns out that the dominant contributions to the resolved part at
large $\eta_{LAB}$ are driven by the polarized photonic {\em gluon} 
distribution $\Delta g^{\gamma}$. From these results it thus 
appears that a measurement of the proton's
$\Delta g$ should be possible from single-jet events at negative rapidities 
where the contamination from the resolved contribution is minimal. 
On the other hand, one can only learn something about the polarized photon 
structure functions if the polarized parton distributions of the
proton are already known to some accuracy. 

In the unpolarized case, an experimental criterion for a distinction  
between direct and resolved contributions 
has been introduced \cite{jeff} and used \cite{jet2ph} in the case of dijet 
photoproduction at HERA. We will now adopt this criterion for the polarized 
case to see whether it would enable a better access to $\Delta g$ and/or the 
polarized photon structure functions. The generic expression 
for the polarized cross section for the photoproduction of two jets with 
laboratory system rapidities $\eta_1$, $\eta_2$ is to LO
\begin{equation} \label{wq2jet}
\frac{d^3 \Delta \sigma}{dp_T d\eta_1 d\eta_2} = 2 p_T 
\sum_{f^e,f^p} x_e \Delta f^e (x_e,M^2) x_p \Delta f^p (x_p,M^2)
\frac{d\Delta \hat{\sigma}}{d\hat{t}} \; ,
\end{equation}
where $p_T$ is the transverse momentum of one of the two jets (which balance
each other in LO) and 
\begin{eqnarray}
x_e &\equiv& \frac{p_T}{2 E_e} \left( e^{-\eta_1} + e^{-\eta_2} \right) \; ,
\nonumber \\
x_p &\equiv& \frac{p_T}{2 E_p} \left( e^{\eta_1} + e^{\eta_2} \right) \; .
\end{eqnarray}
Following \cite{jet2ph}, we will integrate over the cross section to obtain 
$d\Delta \sigma/d\bar{\eta}$, where $\bar{\eta} \equiv (\eta_1 + \eta_2)/2$.
Furthermore, we will apply the cuts \cite{jet2ph}
$$|\Delta \eta| \equiv |\eta_1-\eta_2| \leq 0.5 \; , \;\; 
p_T>6 \; \mbox{GeV} \; \; .$$
The important point is that measurement of the jet rapidities allows 
for fully reconstructing the kinematics of the underlying hard subprocess
and thus for determining the variable \cite{jet2ph}
\begin{equation}
x_{\gamma}^{OBS} = \frac{\sum_{jets} p_T^{jet} e^{-\eta^{jet}}}{2yE_e} \; ,
\end{equation} 
which in LO equals $x_{\gamma}=x_e/y$ with $y$ as before being the 
fraction of the electron's energy taken by the photon. Thus it becomes
possible to experimentally select events at large $x_{\gamma}$, 
$x_{\gamma} > 0.75$ \cite{jeff,jet2ph}, 
hereby extracting the {\em direct} contribution to 
the cross section with just a rather small contamination from resolved 
processes. Conversely, the events with $x_{\gamma}\leq 0.75$ will represent the 
resolved part of the cross section. This procedure should therefore be ideal
to extract $\Delta g$ on the one hand, and examine the polarized photon 
structure functions on the other.

Fig.~9 shows the results for the direct part of the cross section 
according to the above selection criteria. The contributions from the 
resolved subprocesses have been included, using the 'maximally' 
saturated set of polarized photonic parton densities. They turn out
to be non-negligible but, as expected, subdominant. More importantly,
due to the constraint $x_{\gamma}>0.75$ they are determined by the 
polarized quark, in particular the $u$-quark, distributions in the photon, 
which at large $x_{\gamma}$ are equal to their unpolarized counterparts as 
a result of the $Q^2$-evolution (see Fig.~2), rather {\em independent} of 
the hadronic input chosen\footnote{We note that the so-called LO 'asymptotic' 
solutions for the polarized and unpolarized photon structure functions, 
only valid for very large $Q^2$ and $x$, also give $A_f^{\gamma}\rightarrow 1$
as $x\rightarrow 1$.}. Thus the 
uncertainty coming from the polarized photon structure is minimal here 
and under control.
As becomes obvious from Fig.~9, the cross sections are fairly large over the 
whole range of $\bar{\eta}$ displayed and very sensitive to the shape 
{\em and} the size of $\Delta g$ with, however, not too sizeable asymmetries.
A measurement of $\Delta g$ thus could be possible under the imposed 
conditions. Fig.~10 displays the same results, but now for the resolved 
contribution with $x_{\gamma} \leq 0.75$ for the 'maximally' saturated set 
(Figs.~10a,b) and the 'minimally' saturated one (Figs.~10c,d). As expected, 
the results depend on both the parton content of the polarized photon and 
the proton, which implies that the latter has to be known to some accuracy
to extract some information on the polarized photon structure. 
It turns out that again mostly the
polarized {\em gluon} distribution of the photon would be probed in this 
case, in particular at $\bar{\eta}>0.75$. Contributions from the 
$\Delta q^{\gamma}$ are more affected by the $x_{\gamma}$-cut; still such 
contributions amount to about $50\%$ of the cross section at $\bar{\eta}=0$.
We finally emphasize that the experimental finding of a non-vanishing 
asymmetry would establish at least the definite existence of a resolved 
contribution to the polarized cross section. 
\section{Summary and Conclusions}
We have analyzed various conceivable spin-physics experiments at possibly 
forthcoming future polarized $ep$-colliders with high luminosity at HERA and 
the GSI. Here we have studied the charm contribution, $g_1^c$, to the 
polarized DIS structure function $g_1$ and photoproduction of open charm and
jets. All processes we have considered have in common that they get
contributions from incoming gluons already in the lowest order and thus look
promising tools to measure the polarized gluon distribution of the proton. 
In addition, they all have already been successfully studied in the 
unpolarized case at HERA which also provides a guidance concerning the 
experimental cuts to be used in the calculations. 

The DIS charm contribution $g_1^c (x,Q^2=10 \; \mbox{GeV}^2)$ turns out to 
be strongly dependent on the size and shape of $\Delta g$; however, it 
constitutes a sizeable part of $g_1$ only in the region $x<0.05$.

The photoproduction experiments we have studied derive their importance
from their sensitivity not only to $\Delta g$, but also to the completely
unknown parton content of the polarized photon entering via the resolved 
contributions to the polarized cross sections. As far as a 'clear' 
determination of $\Delta g$ is concerned, this resolved piece, if 
non-negligible, might potentially act as an obstructing background, and it is
therefore crucial to assess its possible size. For this purpose, we have 
employed two very different sets for the polarized photonic parton 
distributions which are based on different assumptions concerning the 
hadronic (VMD) input. Conversely, and keeping in mind that HERA has been able
to provide much new information on the unpolarized hadronic structure of the
photon, it is also conceivable that photoproduction experiments at a polarized
version of HERA could be {\em the} place to actually look for effects of the 
polarized photon structure and to prove the existence of resolved
contributions to the polarized cross sections and asymmetries. 

\hspace*{0cm}From these
points of view, our main results can be summarized as follows:
In the case of open-charm photoproduction we found that the resolved 
contribution is generally negligible except for the {\em total} charm cross 
section at HERA energies. Furthermore, the cross sections and their 
asymmetries are very sensitive to shape and size of $\Delta g$. This is true 
in particular for the distribution of the cross section in pseudorapidity,
where also both cross section and asymmetry appear large enough to be 
measurable at HERA and/or the GSI machine with sufficient accuracy to decide 
between the various possible scenarios for $\Delta g$. Concerning 
photoproduction of jets, we find a generally much larger size of the 
resolved contribution. It turns out that the rapidity-distribution of the 
single-inclusive jet cross section separates out the direct part of the 
cross section at negative rapidities. In this region again a strong 
dependence on $\Delta g$ is found with larger cross sections than 
for the case of charm production, but smaller asymmetries. At larger 
rapidities the cross section becomes sensitive to both the parton
content of the polarized proton {\em and} photon, and an extraction 
of either of them does not seem straightforward. The situation improves 
when considering dijet production and adopting an analysis successfully
performed in the unpolarized case \cite{jeff,jet2ph} which is based on 
reconstructing the kinematics of the underlying subprocess and thus 
effectively separating direct from resolved contributions. We find that in 
this case the (experimentally defined) direct contribution should provide 
access to $\Delta g$ whereas the resolved part, if giving rise to
a non-vanishing asymmetry, would establish existence of a polarized 
parton content of the photon. We finally emphasize that the corresponding
measurements will not be easy since the involved asymmetries are
not very large despite sizeable polarized cross sections. With expected 
high luminosities they seem a very interesting challenge for future 
polarized $ep$ colliders at DESY or the GSI.
\section*{Acknowledgements}
We are thankful to M. Gl\"{u}ck for helpful discussions. The work of M.S.\ 
has been supported in part by the 'Bundesministerium f\"{u}r Bildung,
Wissenschaft, Forschung und Technologie', Bonn.

\newpage
\section*{Figure Captions}
\begin{description}
\item[Fig.1] Gluon distributions at $Q^2=10$ GeV$^2$ of the four LO sets of
polarized parton distributions used in this paper. The dotted line refers
to set C of \cite{gs}, whereas the other distributions are taken from 
\cite{grsv} as described in the text. 
\item[Fig.2] Photonic LO parton asymmetries 
$A_f^{\gamma}\equiv \Delta f^{\gamma}/f^{\gamma}$ at $Q^2=30$ GeV$^2$ 
for the two scenarios considered in \cite{gvg,gsvg} (see text). The 
unpolarized LO photonic parton distributions were taken from \cite{grvg}.
\item[Fig.3] {\bf a:} Charm contribution, $g_1^c$, to $g_1$ at 
$Q^2=10$ GeV$^2$ for the four gluon distributions of Fig.~1, calculated 
according to Eq.(\ref{eq1}) using $M=2 m_c$ and $m_c=1.5$ GeV. 
{\bf b:} Charm asymmetry corresponding to {\bf a}. 
\item[Fig.4] Asymmetry for the total charm photoproduction cross section
vs.\ the photon-proton cms energy $\sqrt{S_{\gamma p}}$, calculated 
according to Eq.(\ref{wqctot}) using $M=2 m_c$ and $m_c=1.5$ GeV.
\item[Fig.5] {\bf a:} $p_T$-dependence of the (negative) polarized 
charm-photoproduction 
cross section in $ep$-collisions at HERA, calculated according to 
Eq.(\ref{wqc}) (using $M=m_T/2$ and $m_c=1.5$ GeV) and integrated over 
$-1 \leq \eta_{LAB} \leq 2$. The line drawings are as in the previous figures.
For comparison the resolved contribution to the cross section, calculated
with the 'fitted $\Delta g$' gluon distribution of \cite{grsv} and the 
'maximally' saturated set of polarized photonic parton distributions is
shown by the lower solid line. {\bf b:} Asymmetry corresponding to {\bf a}.
{\bf c,d:} Same as {\bf a,b}, but for the $\eta_{LAB}$ dependence, integrated
over $p_T>8$ GeV.
\item[Fig.6] Same as Fig.~5, but for $E_p=50$ GeV, $E_e=5$ GeV (GSI). For 
{\bf a,b} we have integrated over $-1 \leq \eta_{LAB} \leq 1$ and for {\bf c,d}
over $p_T>3$ GeV. To avoid confusion due to the sign of the cross section, 
the result for set C of \cite{gs} is only shown in the asymmetry plots. 
\item[Fig.7] {\bf a:} $p_T$-dependence of the polarized single-jet inclusive
photoproduction cross section in $ep$-collisions at HERA, integrated over 
$-1\leq \eta_{LAB} \leq 2$. The renormalization/factorization scale was 
chosen to be $M=p_T$. The resolved contribution to the cross section has 
been calculated with the 'maximally' saturated set of polarized photonic 
parton distributions. Note that we show the {\em absolute} value of the 
cross sections; the respective signs can be inferred from {\bf b}.
{\bf b:} Asymmetry corresponding to {\bf a}. {\bf c,d:} Same as {\bf a,b}, 
but for the 'minimally' saturated set of polarized photonic parton 
distributions.
\item[Fig.8] Same as Fig.~7, but for the $\eta_{LAB}$-dependence of the
cross section, integrated over $p_T>8$ GeV.
\item[Fig.9] {\bf a:} $\bar{\eta}$-dependence of the 'direct' part of the 
polarized two-jet photoproduction cross section in $ep$-collisions at HERA 
for the four different sets of polarized parton distributions of the 
proton. The experimental criterion $x_{\gamma}^{OBS}>0.75$ has been applied 
to define the 'direct' contribution (see text). The resolved contribution 
with $x_{\gamma}^{OBS}>0.75$ has been included using the 'maximally' saturated 
set of polarized photonic parton distributions. 
{\bf b:} Asymmetry corresponding to {\bf a}.
\item[Fig.10] Same as Fig.~9, but for the resolved part of the cross section, 
defined by $x_{\gamma}^{OBS}\leq 0.75$ (see text). For {\bf a,b:} the 
'maximally' saturated set of polarized photonic parton distributions has 
been used and for {\bf c,d} the 'minimally' saturated one.
\end{description}
%
\newpage
\pagestyle{empty}

\vspace*{-0.0cm}
\hspace*{-1.4cm}
\epsfig{file=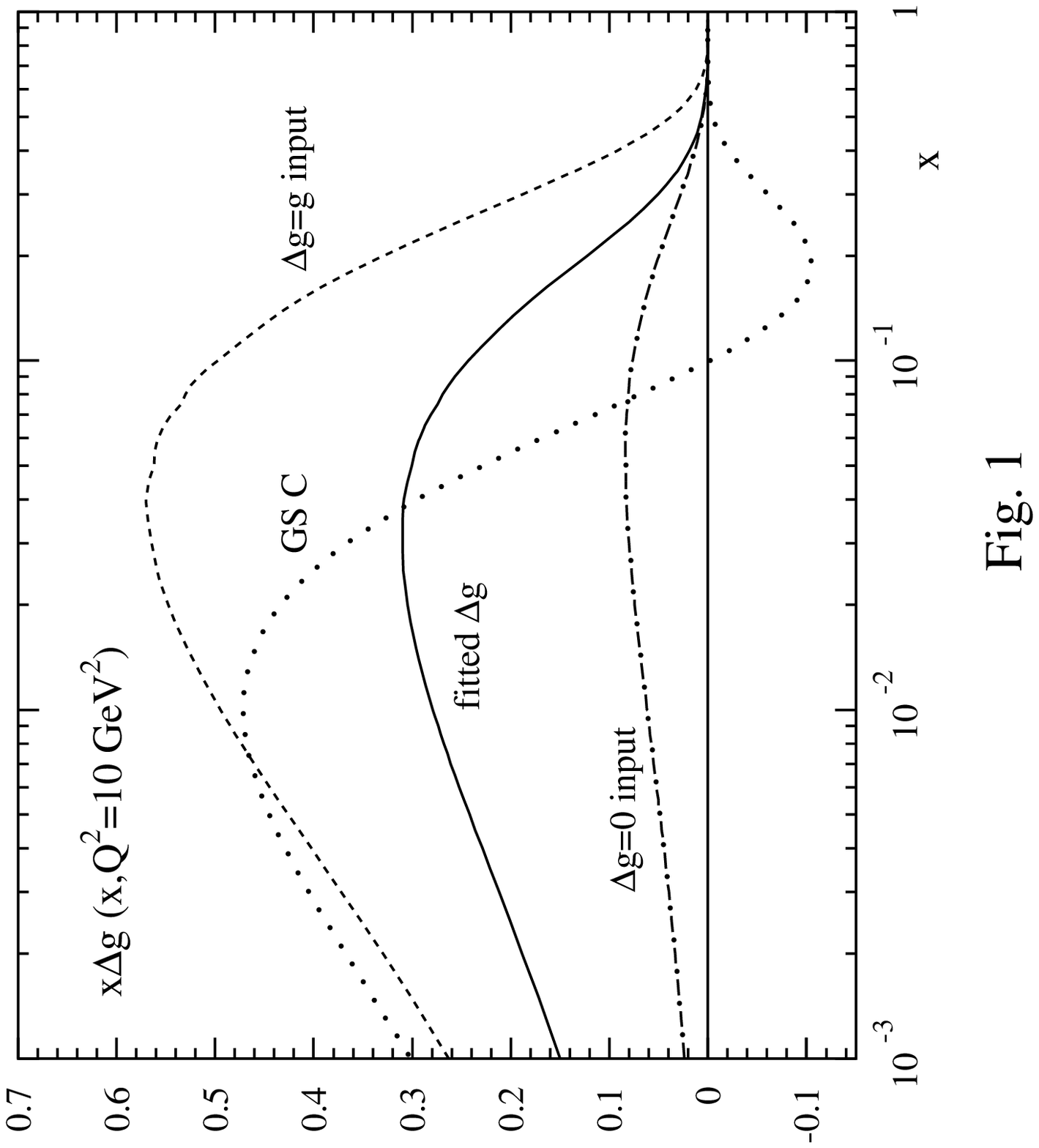,angle=90}
\newpage

\vspace*{-1.0cm}
\hspace*{-1.4cm}
\epsfig{file=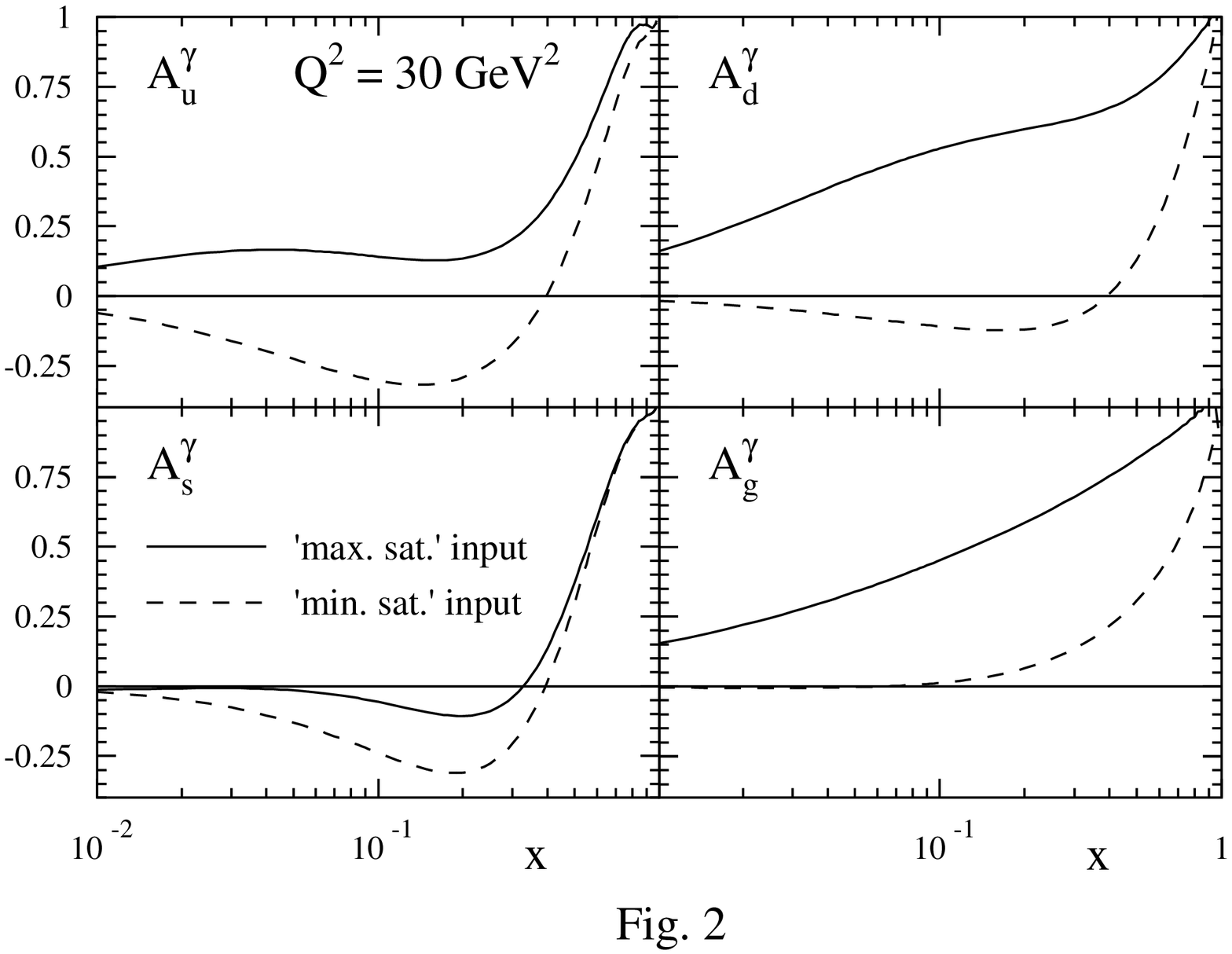,angle=90}
\newpage

\vspace*{-0.5cm}
\hspace*{-0.6cm}
\epsfig{file=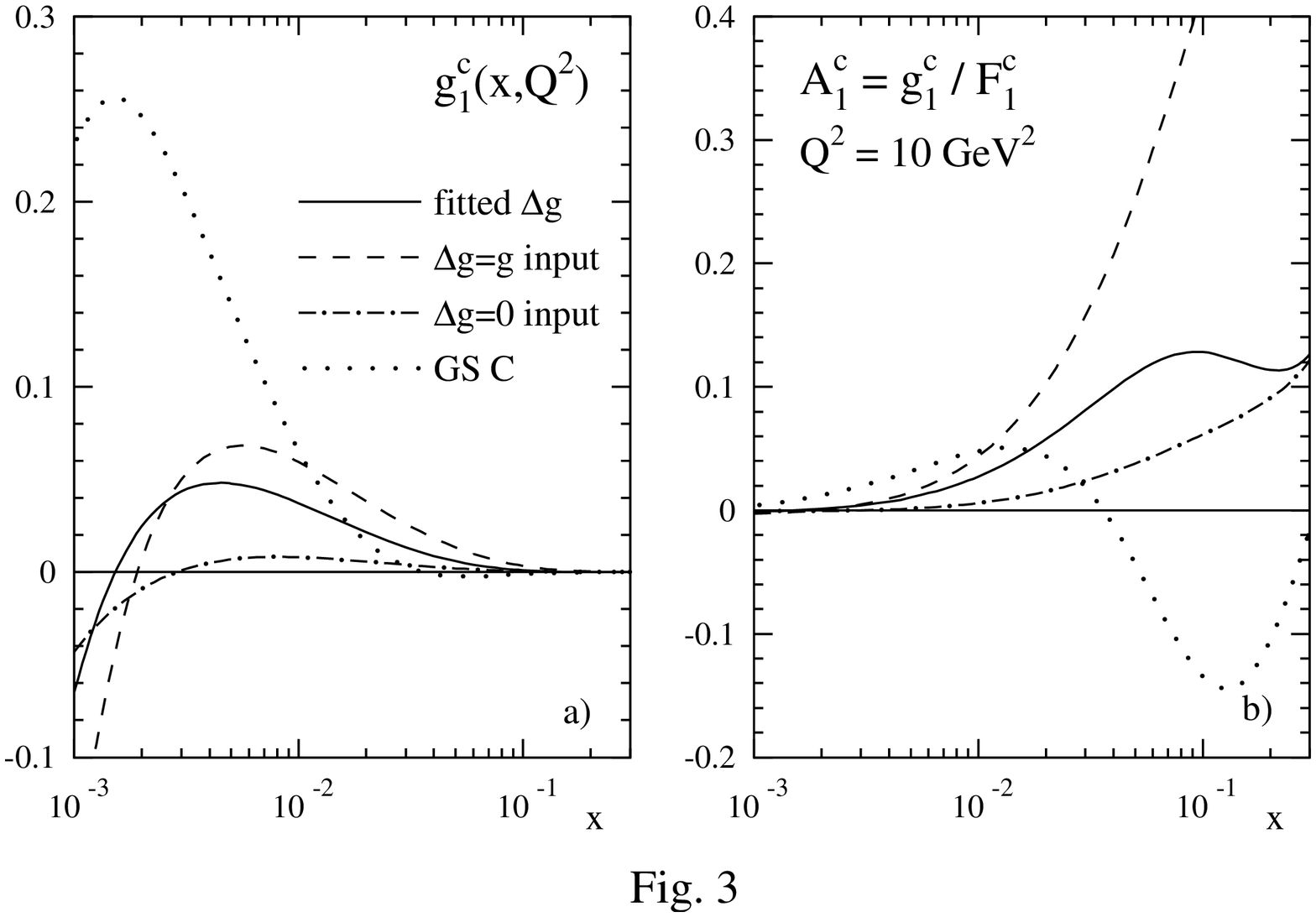,angle=90}
\newpage

\vspace*{-0.0cm}
\hspace*{-1.4cm}
\epsfig{file=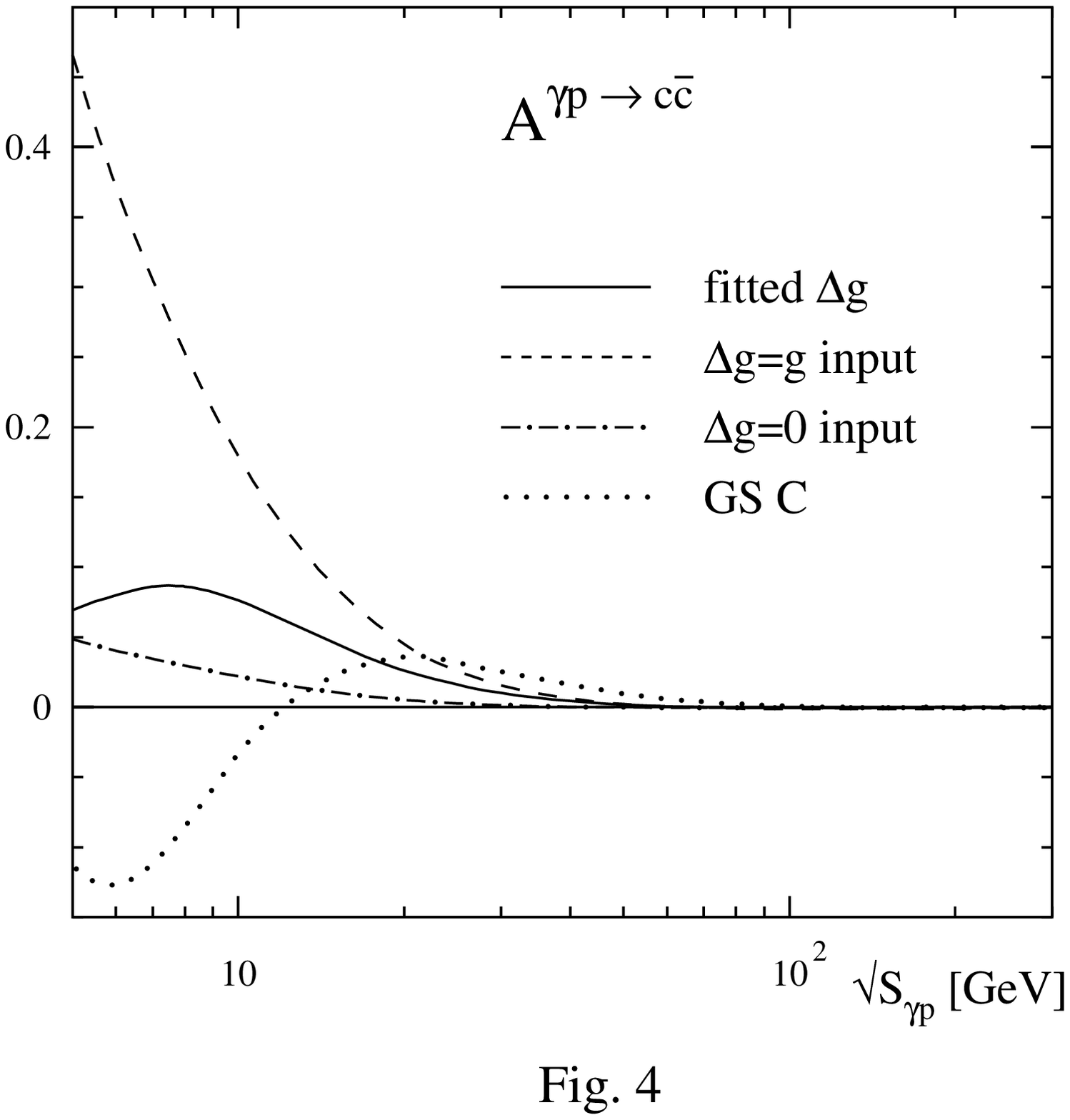,angle=0}
\newpage

\vspace*{-0.5cm}
\hspace*{-2.4cm}
\epsfig{file=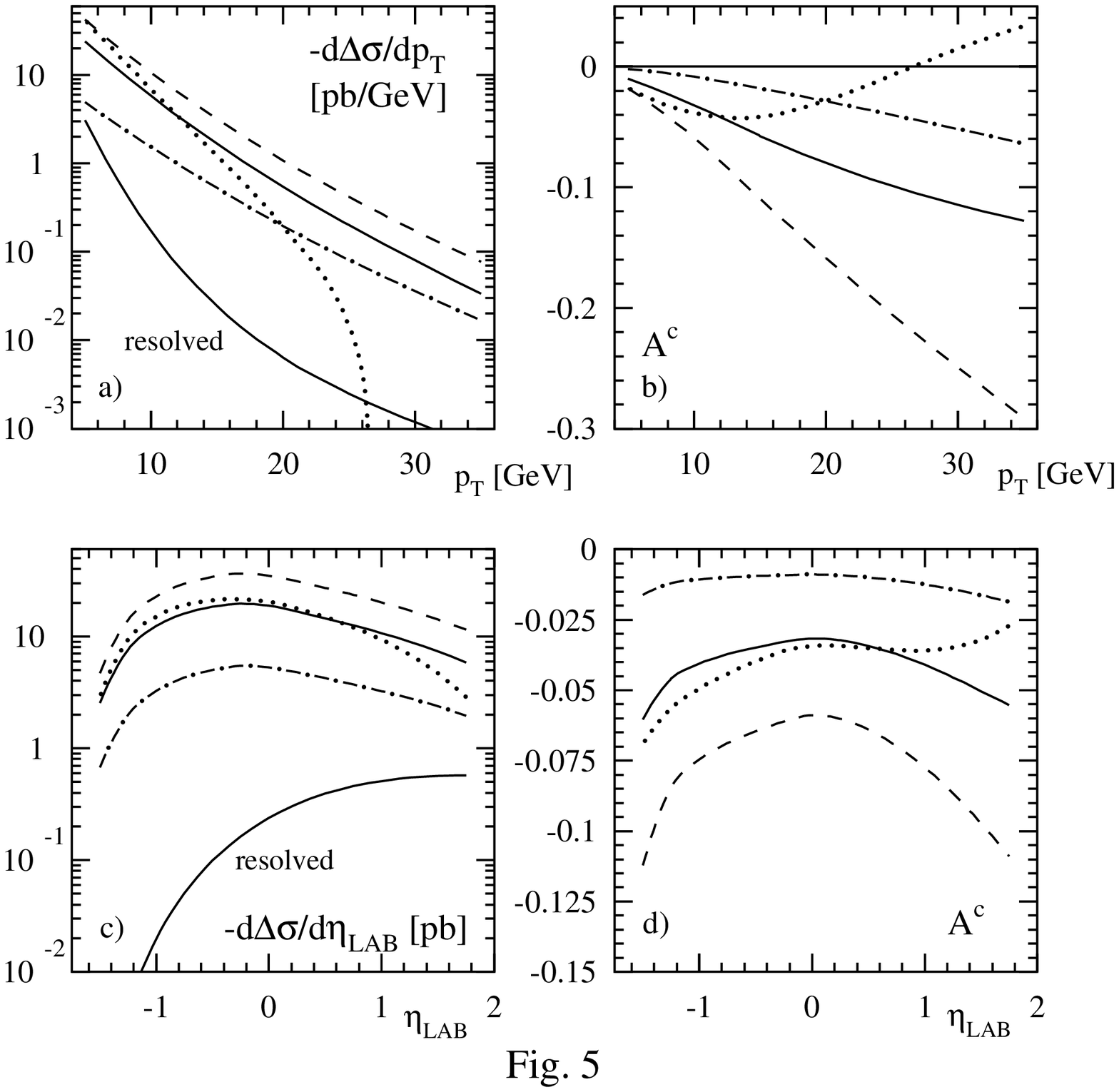,angle=0}
\newpage

\vspace*{-0.5cm}
\hspace*{-2.4cm}
\epsfig{file=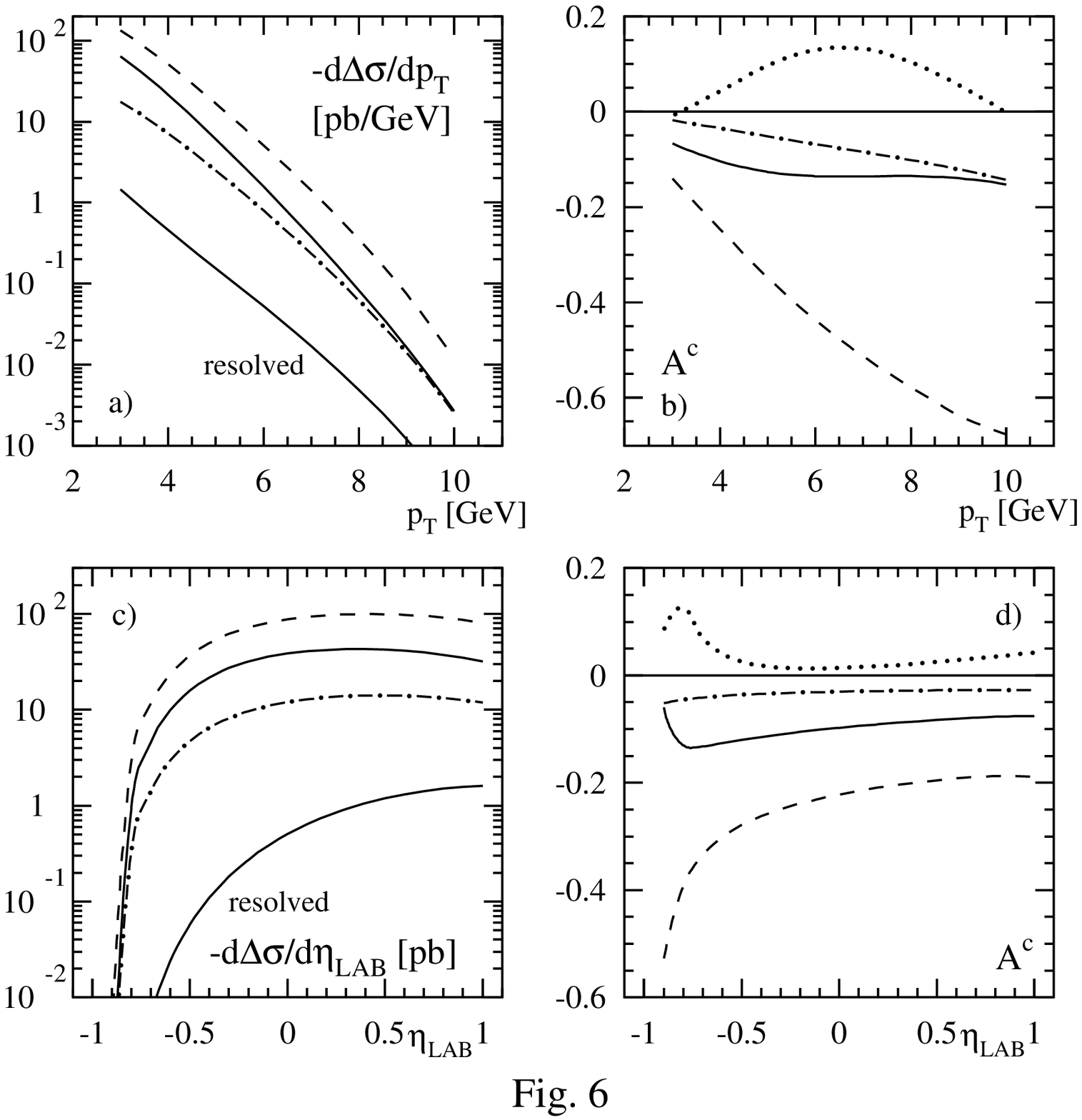,angle=0}
\newpage

\vspace*{-1.0cm}
\hspace*{-2.4cm}
\epsfig{file=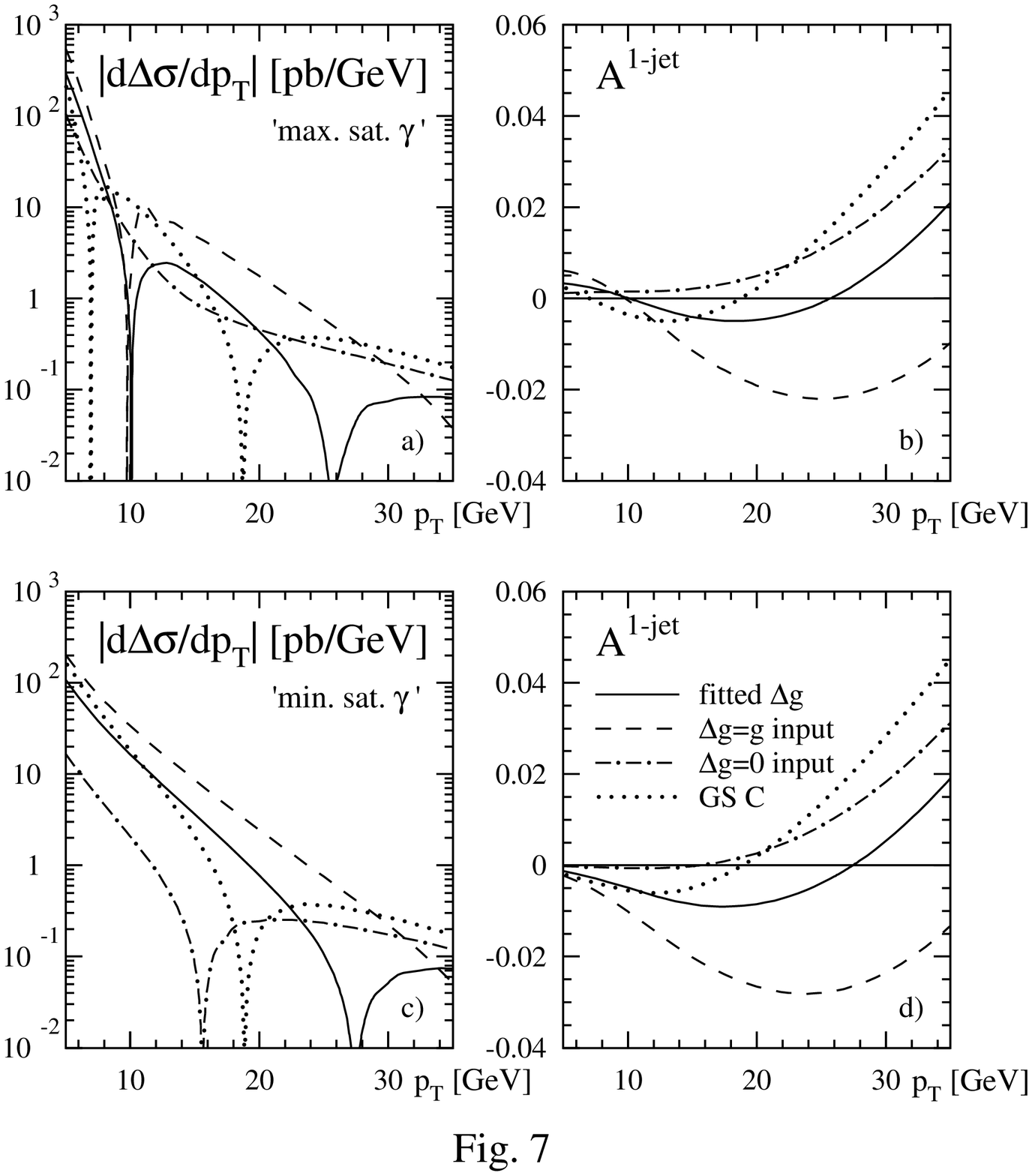,angle=0}
\newpage

\vspace*{-1.0cm}
\hspace*{-2.4cm}
\epsfig{file=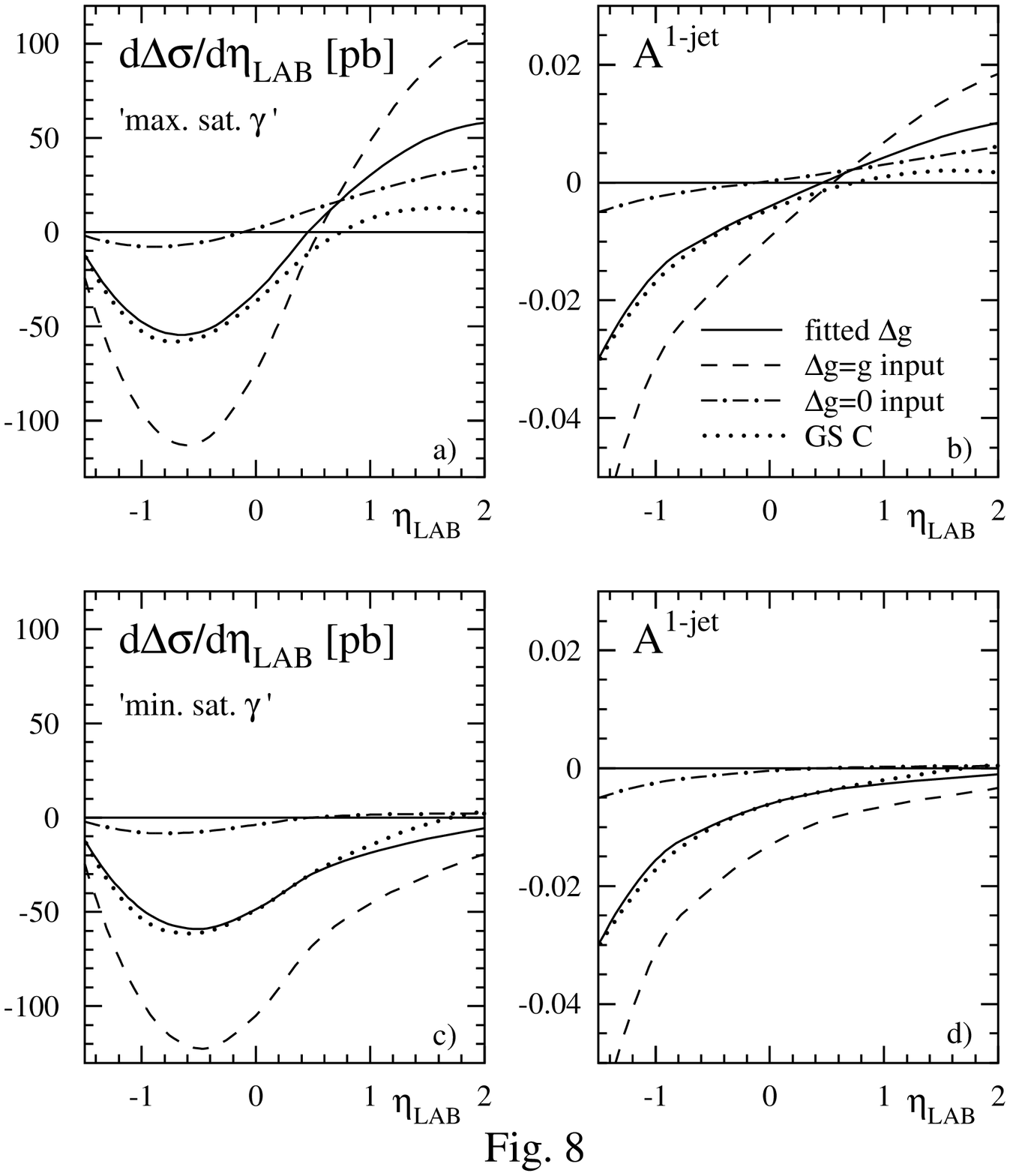,angle=0}
\newpage

\vspace*{-0.5cm}
\hspace*{-1.4cm}
\epsfig{file=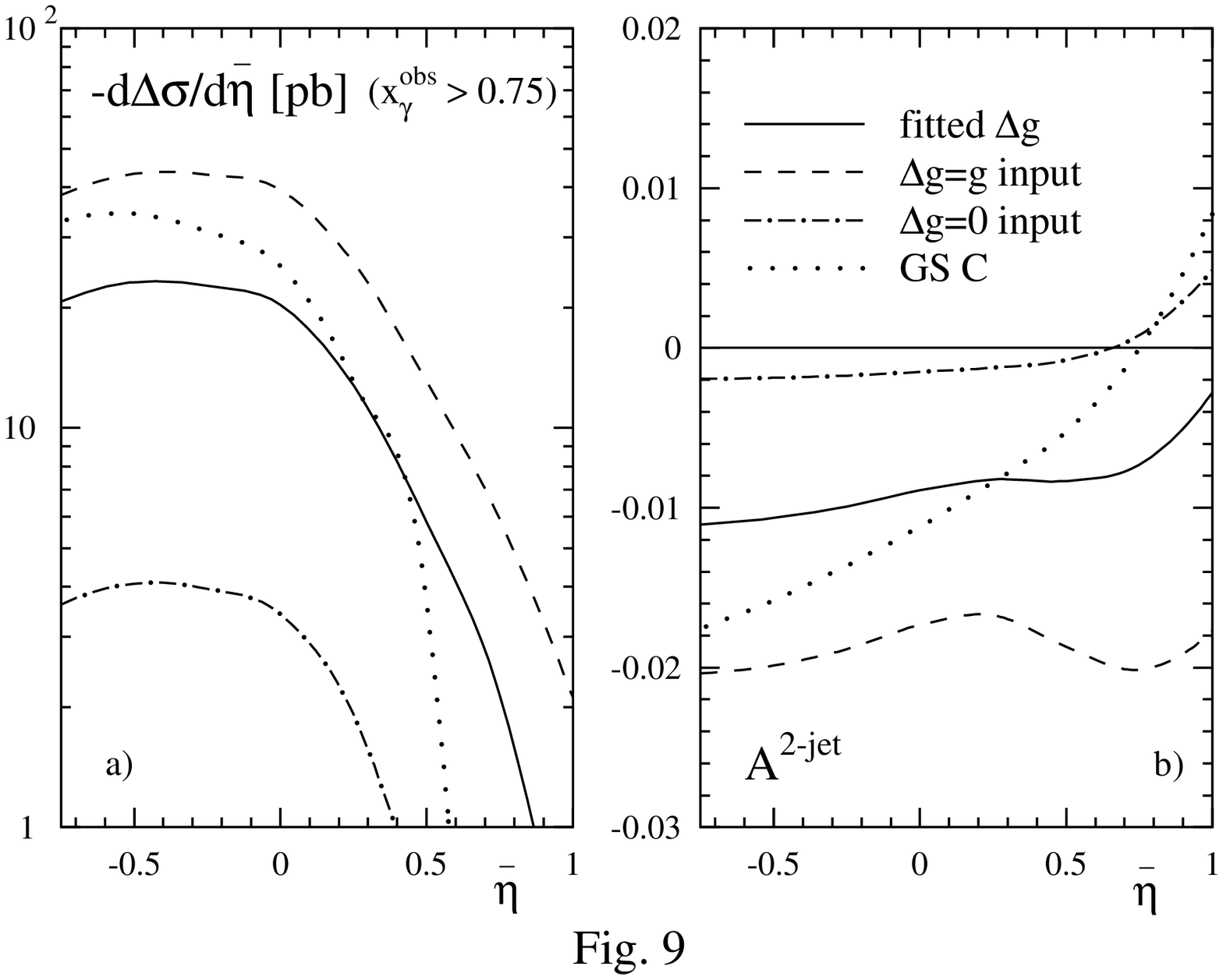,angle=90}
\newpage

\vspace*{-0.8cm}
\hspace*{-2.3cm}
\epsfig{file=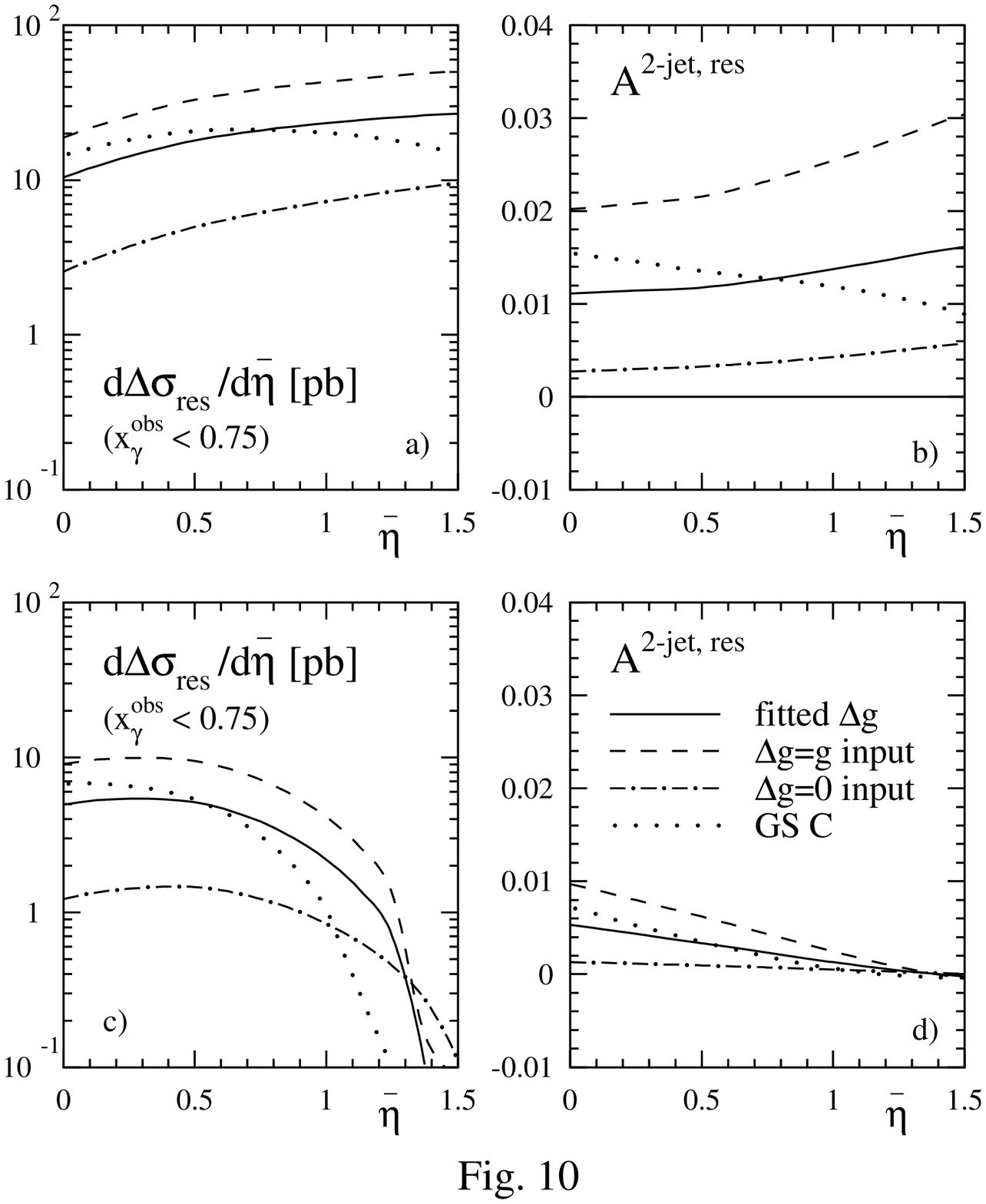,angle=0}
\end{document}